\documentclass[a4paper,twocolumn,]{emulateapj}

\usepackage{epsfig,wrapfig,color,sidecap}
\usepackage{index}
\usepackage{subfigure}
\usepackage{mathtools}

\usepackage{colordvi}

\usepackage[usenames,dvipsnames]{xcolor}
\usepackage[normalem]{ulem}

\usepackage{times}
\usepackage[utf8x]{inputenc}
\usepackage{mathrsfs}

\usepackage{multirow}

\shorttitle{}
\shortauthors{Benyamin et al.}

\RequirePackage[colorlinks=true
,urlcolor=blue
,anchorcolor=blue
,citecolor=blue
,filecolor=blue
,linkcolor=blue
,menucolor=blue
,pagecolor=blue
,linktocpage=true
,pdfproducer=medialab
]{hyperref}

\begin{document}

\bibliographystyle{authordate1}

\title{Electron-capture isotopes could constrain cosmic-ray propagation models}

\author{David Benyamin, Nir J. Shaviv \& Tsvi Piran} 
\affil{The Racah Institute of physics, The Hebrew University of Jerusalem, Jerusalem 91904, Israel}

\begin{abstract}

Electron capture (EC) isotopes are known to provide constraints on the low energy behavior of cosmic rays (CRs), such as re-acceleration. Here we study the EC isotopes within the framework of the dynamic spiral-arms CR propagation model in which most of the CR sources reside in the galactic spiral arms. The model was previously used to explain the B/C and sub-Fe/Fe ratios \citep{BoverC,Iron}. We show that the known inconsistency between the $^{49}$Ti/$^{49}$V and $^{51}$V/$^{51}$Cr ratios remains also in the spiral-arms model. On the other hand, unlike the general wisdom in which the isotope ratios depend primarily on reacceleration, we find here that the ratio also depends on the halo size ($Z_\mathrm{h}$) and in spiral-arms models also on the time since the  last spiral arm passage ($\tau_\mathrm{arm}$). Namely, EC isotopes can in principle provide interesting constraints on the diffusion geometry.  However, with the present uncertainties in the lab measurements of both the electron attachment rate and the fragmentation cross-sections, no meaningful constraint can be placed.
\end{abstract}


\section{introduction}
\label{sec:intro}
\maketitle

Observations of the CR composition can teach about the origin of CRs, their initial composition, path length distribution (PLD) and interaction they undergo as they propagate in the interstellar medium. Such observations include the ratio between secondary to primary cosmic rays, including the Boron to Carbon (B/C) ratio and the sub-Iron (Scandium through Manganese) to Iron (sub-Fe/Fe) ratio, the positron fraction ($e^{+}/(e^{-}+e^{+})$), and the ratios between the EC's  daughter and parent isotopes which we study here. The latter are known to constrain the process of re-acceleration, but as we show below, they can also be used, at least in principle, to constrain diffusion models. 

If one calculates the secondary to primary ratio under the simplest leaky box model or a ``disk-like" model (with an azimuthally symmetric CRs source distribution), one finds that the ratio drops with energy \citep{CRreview}. Indeed, the positron fraction below 10 GeV and the nuclei ratios above 1 GeV/nuc.\ exhibit this behavior. However, the positron fraction above 10 GeV and nuclei spectra below 1 GeV/nuc.\ appear to be behave differently \citep{Adriani,AMS-preliminary,StrongNucleons}. A disk-like model must include galactic winds, re-acceleration or ad-hoc assumptions on the diffusivity in order to explain the observed behavior of the nuclei ratios \citep{StrongReview,Dragon}, while the behavior of the positrons require either astrophysics solutions such as pulsars \citep[e.g.,][]{Pulsar1,Pulsar2,Pulsar3,Pulsar4,Pulsar5}, or more exotic physics such as dark matter decay \citep[e.g.,][]{DarkMetter1,DarkMetter2}

Unlike a disk-like model, a spiral-arms model, in which  a significant fraction of CR sources are located at the galactic spiral arms---a place where star formation is enhanced and hence young SNRs are abundant, can explain these anomalies at the outset. \cite{Pamela} showed that by considering the CR sources to be at a finite distance from earth, as expected from the spiral structure, one recovers the positron fraction spectrum. Moreover, \cite{BoverC} recovered the B/C ratio also at low energies by taking the spiral arms to be dynamic.  The astrophysical motivation of the spiral arms model and its success  in explaining these phenomena has motivated us to explore other predictions of this model.  

Another interesting problem arises when the grammage required to explain the B/C ratio is compared to the sub-Fe/Fe ratio. When doing so, it reveals that the latter ratio requires about 20\% more grammage than the former ratio in a disk-like model \citep{GarciaMunoz,Davis}. \cite{GarciaMunoz} proposed a solution to this problem, by cutting the short path lengths from the CR PLD. Obviously, taking the source to be primarily in the spiral arms naturally causes a paucity in short path lengths. In a later study  the Ulysses-HET group \citep{duvernois1996interstellar}  also tested the option of truncating the short path lengths.  They found that the this is not necessary. Namely, a simple exponential power-law (as is the case in a disk-like and leaky-box models) is sufficient to recover their own observations. However, the Ulysses-HET sub-Iron/Iron measurement is clearly well below other measurement (while the B/C is consistently the same), which explains why a simple PLD is enough when only their measurement is considered.
In \cite{Iron}, we have shown that  a spiral-arms model resolves this anomaly by finding the optimal model parameters required to separately recover the B/C and sub-Iron/Iron. It was shown that while the disk-like model does not recover the two ratios with consistently the same model parameters, the spiral arm model does. 

Here we continue our investigation of the Iron group CR nuclei (Scandium through Nickel)\footnote{The iron group includes the isotopes of Scanadium through Nickel, with Iron and Nickel being the primaries and the rest secondaries. the sub-Iron group is a sub group of the iron group that includes Scandium through Manganese.} within the context of the spiral arm model, and focus on isotopes that decay through EC. At low energies, electrons are bound to the nuclei and as a consequence, these isotopes rapidly decay through EC. However, at higher energies, typically above 1 GeV/nuc., these isotopes are stripped of their electrons and this inhibits their decay. The probability for having bound electrons depends  on two processes, the stripping and the attachment of electrons. Both  are strongly dependent on energy \citep{Letaw}. However, because the EC decay time scale is generally much shorter than the stripping time scale, the isotope ratios basically depends on the attachment rate. 

\cite{ACE} reported the first measurements of $^{44}$Ti, $^{49}$V, $^{51}$Cr, $^{55}$Fe and $^{57}$Co from CRIS -- the Cosmic Ray Isotope Spectrometer, that is located on the Advanced Composition Explorer (ACE). In particular, they measured the $^{49}$Ti/$^{49}$V and $^{51}$V/$^{51}$Cr ratios. Evident from the observations is a very strong dependance on energy---the ratios decrease with energy. Since the enumerator in these ratios is the stable daughter product of the EC of the parent isotope in the denominator, the ratios reflect the EC reaction rate, and therefore the probability for the parent isotope to be stripped. 

\cite{Jones} modeled the isotopic ratios using the weighted slab model, while assuming different assumptions on the retainment of electrons and reacceleration. They have shown that complete stripping results in almost energy independent ratios, and therefore cannot explain the decrease with energy.  In other words, there must be a transition from unstripped to stripped isotopes.  The decrease with energy of the two ratios $^{49}$Ti/$^{49}$V and $^{51}$V/$^{51}$Cr is consistent with this interpretation as well. 
    
\cite{Jones} then tried to explain the measurements by assuming that the nuclei retain their bound electrons at low energies, and then reaccelerate to higher energies, on their way to the solar system. The time spent at low energies will cause the EC isotopes to produce more daughter isotopes compared to CRs that did not spend time at low energies. 

In a subsequent study, \cite{Niebur} showed that the cross-section to bind electrons from the ISM to stripped nuclei  is increasing for progressively smaller energies. For energies lower than a few hundred MeV/nuc., the time scale is shorter than the escape (and therefore typical age) of the cosmic rays. However, even at a few MeV/nuc.\ the attachment rate time scale is still much longer than the EC decay.  This means that at energies of up to a few 100 MeV/nuc., the attachment process is the dominant one determining the EC isotope ratios. They also considered reaccelaration as \cite{Jones}, but due to the large electron attachment cross-section which they include,  \cite{Niebur} require a more feasible higher initial energy to accelerate these isotopes from than \cite{Jones} require. 

However, both \cite{Jones}, \cite{Niebur} obtained inconclusive results---some of the observations were more consistent with models that include reacceleration (in particular, the $^{51}$V/$^{51}$Cr isotopes ratios) while other observations indicate the opposite (the $^{49}$Ti/$^{49}$V isotopes ratio). Both \cite{Jones} and \cite{Niebur} point out that the main problem in reaching any firm conclusions was the uncertainty in the fragmentation cross sections. Namely, the above inconsistency cannot be resolved with just reacceleration. 

This conclusion about the fragmentation cross-sections was reaffirmed by \cite{Niebur2003}, who showed that the typical 10-20\% uncertainty in the fragmentation cross-sections \citep{WebberSoutoulcross}, can explain away the discrepancy between the above two isotope datasets. For example, reducing the $^{49}$Ti fragmentation cross-section by 15\% will resolve the discrepancy. We elaborate on this correction in the discussion. 

We note that the CRIS/ACE results are not the first to have reached these conclusions. They are consistent with the previous measurements by the Ulysses HET team \citep{connell1999ulysses} of a single data point at 300\,MeV/nuc. (but with a similar error bar). These authors  also concluded that some isotopes are consistent with reacceleration and while others are consistent with no acceleration.

All these studies were done within the standard disk model. One could have hoped that like other inconsistencies,  this one will be resolved when we consider a dynamical spiral arms model instead. 
Here we show that the inconsistency between atomic mass 49 and 51 isotopes remains also in this model.
This points out to the same conclusion that there might be a problem with the fragmentation cross-sections.
We also show that while the power law index of the cross-sections' energy dependence required to fit the observations agree with the lab experiments \citep{Wilson,Crawford}, the attachment normalization needed to fit the data varies depending on the halo size,  $z_\mathrm{h}$, and the time since last spiral arm passage, $\tau_\mathrm{arm}$. 

We begin in \S\ref{sec:model} by briefly describing the spiral arms model and the  nominal model parameters. We review the data used in \S\ref{sec:data}. In \S\ref{sec:results} we carry out an extensive analysis of the model, including a parameter study used to find a fitting formula for the attachment rate which recovers the $^{49}$Ti/$^{49}$V and $^{51}$V/$^{51}$Cr ratios. 
The implications of these results are discussed in \S\ref{sec:discussion}.

\section{The model}
\label{sec:model}
 
SNRs are generally believed to be the sources of the galactic CRs. The  spiral-arms model assumes that, since SNR are more abundant in galactic spiral arms, these arms are    the main source of CRs. At low energies the CRs diffuse slowly and the dynamical motion of the spiral arms cannot be neglected. 
In \cite{BoverC} we describe a fully three dimensional numerical code for  CRs diffusion in the Milky Way under these assumptions.  The code enables us to explore 
 {\em dynamic} spiral arms as the main source of the CR. Using this  model, \cite{BoverC} recovered the B/C ratio and demonstrated that the dynamics of the spiral arms has a notable effect on the ratio between secondary and primary CRs, which below 1\,GeV/nuc.\ increase with the energy.

In \cite{Iron} we have shown  that a spiral-arms model, unlike a disk-like model, can explain the discrepancy between the grammage implied by  the B/C ratio and by the sub-Fe/Fe ratio. 
 Naturally, the spiral arms model require different diffusion parameters than those commonly used in the galactic disk model. 
The optimal parameters required to fit the B/C, sub-Iron/Iron and $^{10}$Be/$^{9}$Be ratios within the dynamic arms and homogeneous disk models are summarized in table\ \ref{table:parameters} \citep{Iron, BoverC}. 

\begin{table}[h]
\centering
\caption{Nominal Model Parameters}
\begin{tabular}{ c c c c c c c c}
\hline
 \multirow{3}{*}{parameter} & \multirow{3}{*}{Definition} & value for & value for\\
   & & spiral arm & disk-like\\
   & & model & model\\
\hline
\vspace*{2mm}
$Z_\mathrm{h}$ & Half halo height & 250 ~pc & 1~kpc\\
$D_0$ 
& Diffusion coefficient & $1.2 \times 10^{27}$ 
& $5 \times 10^{27}$
\\
\vspace*{2mm}
 & normalization 
 \footnote{We assume a power law dependent diffusion,  $D = D_0 \beta (R/3$~GV$)^\delta$, where $R$ is the rigidity and $\beta$ is $v/c$.  However, throughout the paper, the term ``diffusion coefficient" actually refers to the normalization $D_\mathrm{0}$ and not $D(R)$.} &  ~cm$^2$/sec &  ~cm$^2$/sec\footnote{Note that the disk-like model diffusion coefficient is at the lower end of values found in the literature. This is because the model has a relatively small halo (1 kpc), but also because we require the model to recover the sub-iron/iron ratio and not the B/C ratio typically fitted in the literature.}  \\
 \vspace*{2mm}
$\delta$ & Spectral index & 0.4 
& 0.5
\\
\vspace*{2mm}
$\tau_\mathrm{arm}$ & Last spiral arm passage & 5 ~Myr & \\ 
\vspace*{2mm}
$i_4$ & 4-arms set's pitch angle & 28$^\circ$ & \\
\vspace*{2mm}
$i_2$ & 2-arms set's pitch angle & $11^{\circ}$ & \\
\multirow{2}{*}{$\Omega_4$} & Angular velocity of & \multirow{2}{*}{15 (km/s) kpc$^{-1}$} & \multirow{2}{*}{} \\
\vspace*{2mm}
 & the 4-arms set & & \\
\multirow{2}{*}{$\Omega_2$} & Angular velocity of & \multirow{2}{*}{25 (km/s) kpc$^{-1}$} & \multirow{2}{*}{} \\
\vspace*{2mm}
 & the 2-arms set & & \\
\multirow{2}{*}{$f_\mathrm{SN,4}$} & Percentage of SN in & \multirow{2}{*}{48.4\%} & \multirow{2}{*}{} \\
\vspace*{2mm}
 & the 4-arms set & & \\
\multirow{2}{*}{$f_\mathrm{SN,2}$} & Percentage of SN in & \multirow{2}{*}{24.2\%} & \multirow{2}{*}{} \\
\vspace*{2mm}
 & the 2-arms set & & \\
\multirow{2}{*}{$f_\mathrm{SN,CC}$} & Percentage of core collapse & \multirow{2}{*}{8.1\%} & \multirow{2}{*}{80.7\%} \\
\vspace*{2mm}
 & SNe in the disk & & \\
\multirow{2}{*}{$f_\mathrm{SN,Ia}$} & Percentage of  & \multirow{2}{*}{19.3\%} & \multirow{2}{*}{19.3\%} \\
 & SN Type Ia & & \\
\hline
\end{tabular}
\label{table:parameters}
\end{table}

Our code is different from present day simulations (such as {\sc galprop}, \citealt{StrongNucleons}, and {\sc dragon}, \citealt{Dragon}) which solve the partial differential equations (PDE) describing diffusion in that we use a Monte Carlo methodology. It allows for more flexibility in adding various physical aspects to the code (such as the spiral arm advection), though at the price of reduced speed. Here we will only discuss the changes we recently made to explore the EC reactions. The full details of the code and of the the model are found in \cite{BoverC,Iron}.

\subsection{Attachment Rate Formula}
\label{sec:formula}

\cite{Letaw} studied the EC reaction in CRs using experimental data collected by \cite{Wilson} and \cite{Crawford}. In figs.~1 and 2 of \cite{Letaw}, one can see that for $21<Z<28$ and for energies of a few 100~MeV/nuc.\ the mean free path for attachment of an electron is roughly $\lambda_\mathrm{attachment} \approx 1~$gr/cm$^2$, while for the stripping of an electron it is roughly $\lambda_\mathrm{stripping} \approx 10^{-3} $\,gr/cm$^2$, which correspond to time scales of $\tau_\mathrm{attachment} \approx 5~$Myr and  $\tau_\mathrm{stripping} \approx 5 \times 10^{-3}$~Myr respectively.  For the isotopes $^{44}$Ti, $^{49}$V, $^{51}$Cr, $^{55}$Fe and $^{57}$Co the decay time scale is between several days to a few years, much smaller than $\tau_\mathrm{stripping}$, implying that we can neglect the stripping process for those isotopes and assume that they decay immediately after they attach an electron from the ISM. However for $^{53}$Mn and $^{59}$Ni, the half life time for the EC decay is 3.7\,Myr and 0.076\,Myr respectively, which is much longer than  $\tau_\mathrm{stripping}$. This allows one to neglect the decay process and assume that these isotopes will be striped off their electrons before they could decay, and therefore remain stable.\footnote{We note that $^{54}$Mn is also an EC isotope. In our calculations it decay immediately since its $\beta$ decay mode have half life time that is significantly shorter than the typical propagation time.}

When interpolating the data of \cite{Letaw}'s fit for the electron attachment mean free path, one can see that the energy dependance of the attachment cross-section is a power-law of the form 
\begin{equation} 
\sigma_{a}(E,Z)=N Z^{\nu} {(E/\mathrm{500\,MeV})^{-\mu}} \ ,
\end{equation}  with  indices of $\mu=1.8 \pm 0.1$ and $\nu=4.5 \pm 0.1$, and a normalization $N=(1.2 \pm 0.2) \times 10^{-4}~$mb for $20<Z<28$.

Here we allow for a generalized power-law attachment rate and add it to the description of EC isotopes in the numerical code, which includes $^{44}$Ti, $^{49}$V, $^{51}$Cr, $^{55}$Fe and $^{57}$Co. Namely,  we use the same power-law with the above two parameters, the normalization factor, $N$, and the power index, $\mu$, but keep them as free parameters (we choose to keep $\nu$ fixed because the difference between the Z's of the two observational datasets is less than $5\%$, see \S\ref{sec:data}). Each time step we check whether the CR isotope attached an electron from the ISM (and let it decay immediately) with the same methodology as we do for the spallation process (see more details in \citealt{BoverC}, \S3.10). In this work we study the sensitivity of isotope ratio outcome to the parameter space describing the attachment. The observed ratios which we use are  $^{49}$Ti/$^{49}$V and $^{51}$V/$^{51}$Cr described below in \S\ref{sec:data}. 

\section{Observational datasets}
\label{sec:data}

We compare the model predictions for the $^{49}$Ti/$^{49}$V and $^{51}$V/$^{51}$Cr ratios with the two CRIS datasets \citep{Niebur2003}, one collected during the solar minimum years, 1997-1999, and one for the solar maximum years, 2000-2003, with average solar modulation of 510~MV and 920~MV respectively \citep{UsoskinPhi}. For each observation, there are 14 data points between 100 MeV/nuc. and 1 GeV/nuc.

To account for the solar wind modulation, the energy of each specie outside the solar system obtained in the simulation is mapped to the modulated energy inside the solar system through $E_{obs}=E-(Z/A) \times \phi$, where $\phi$ is the modulation potential. The modeled specie ratios can then be calculated and compared with the observations at a given observed energy from which a $\chi^2$ can be calculated. 

\begin{figure*}
\centerline{\includegraphics[width=3.0in]{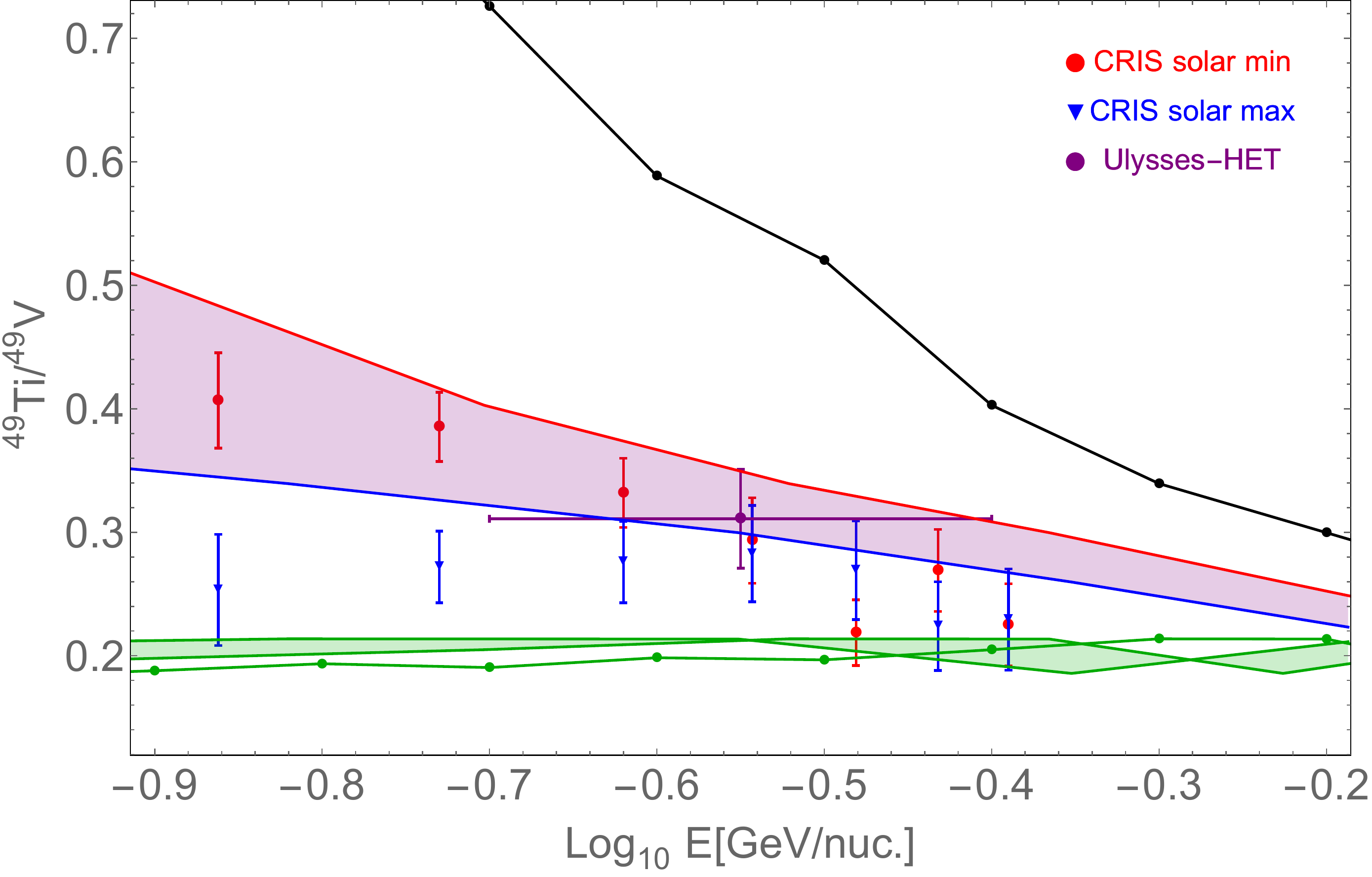} \includegraphics[width=3.0in]{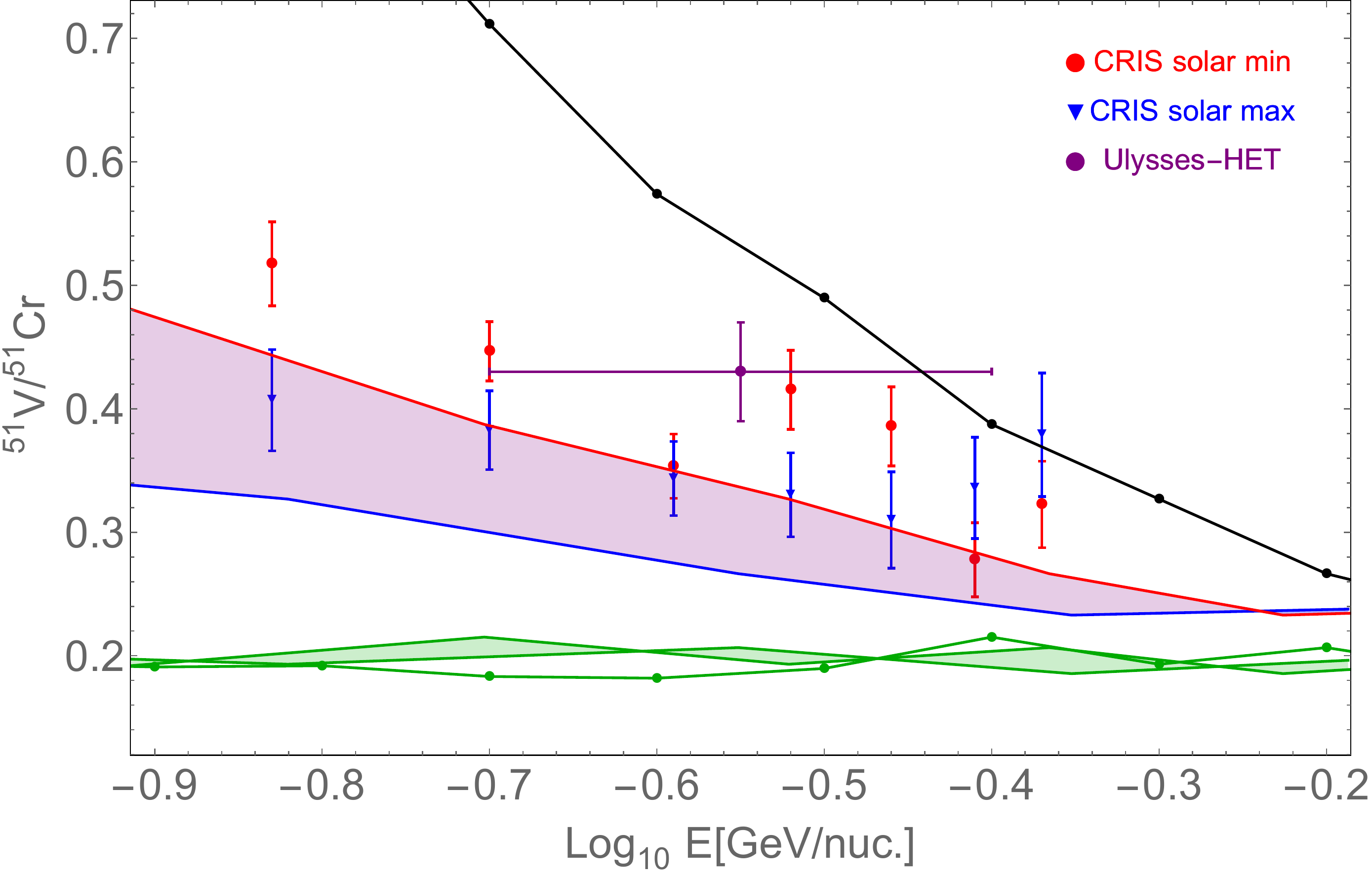}}
\caption{The $^{49}$Ti/$^{49}$V (left figure) and $^{51}$V/$^{51}$Cr (right figure) ratios we obtain in our nominal model (black lines) with the set of parameters described in table\,\ref{table:parameters} and the attachment parameters obtained from the fit described in \S \ref{sec:fit}. The shaded regions correspond to the spectrum once solar wind modulation is added, with the red lines describing the minimum solar modulation while the blue describe the maximum solar modulation. The green lines are the respective lines obtained when the EC isotopes are assumed to be entirely stable. Data taken from: CRIS \citep{Niebur2003}, Ulysses-HET \citep{connell1999ulysses}.}
\label{fig:K}
\end{figure*}

The data (and the model fits) are depicted in figs.\ \ref{fig:K}.

\section{Results}
\label{sec:results}

\subsection{A disk-like model}
\label{sec:disk model}

\begin{figure}
\centerline{\includegraphics[width=3.0in]{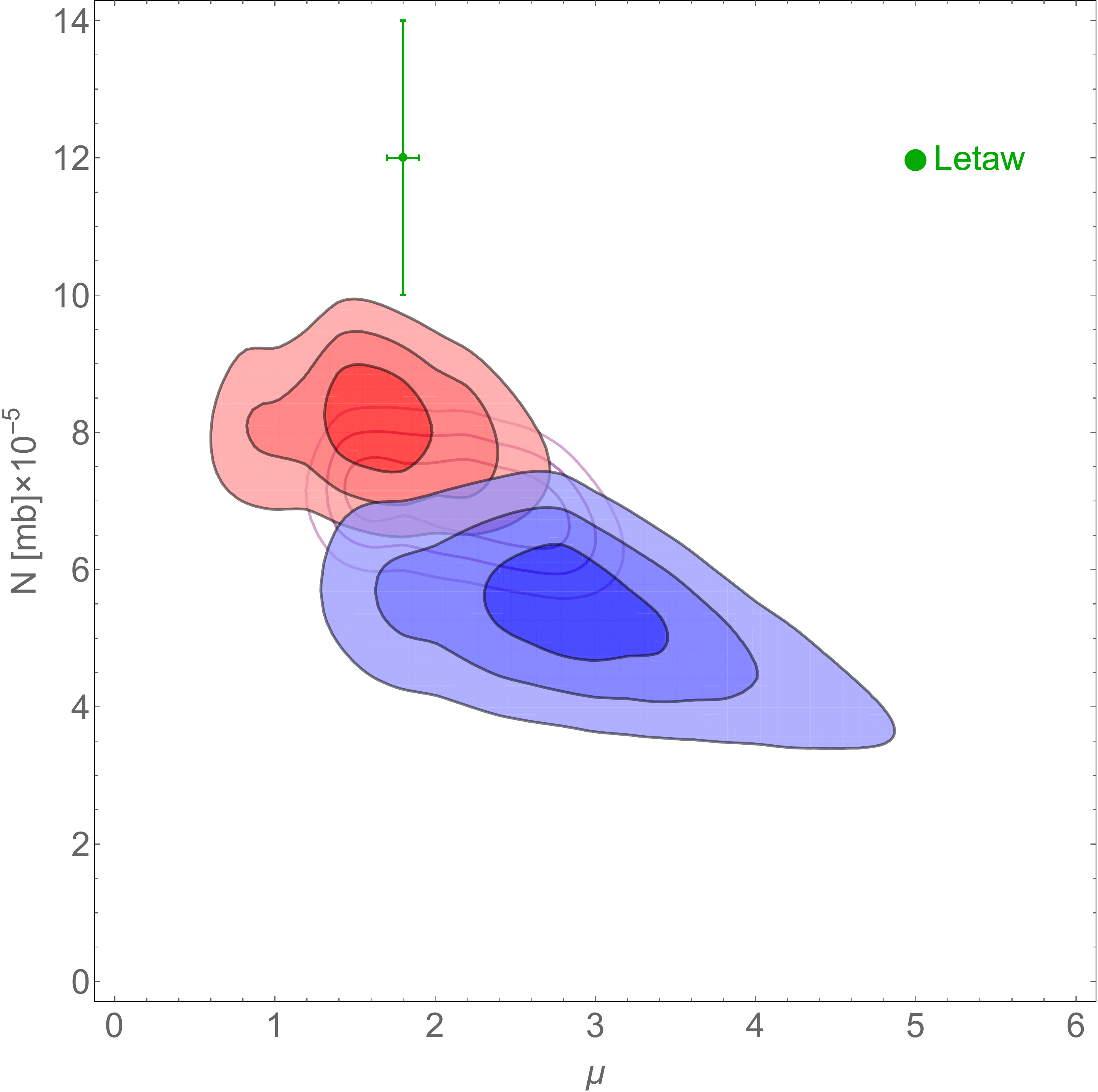}}
\caption{A contour plot of $\chi^2$ fit for a disk-like model with halo size of $z_h=1$~kpc and diffusion coefficient of D$_0=4 \times 10^{27}$~cm$^{2}/$sec. The red contours correspond to the $^{51}$V/$^{51}$Cr fit, the blue contours correspond to the $^{49}$Ti/$^{49}$V fit and the dashed purple lines correspond to the combined $\chi^2$ calculation. Note the discrepancy between the $^{51}$V/$^{51}$Cr and the $^{49}$Ti/$^{49}$V fits---while the observation for $^{51}$V/$^{51}$Cr require high normalization and a low power-law index, the observation for $^{49}$Ti/$^{49}$V require a smaller normalization and a higher power-law index. The green point denotes the electron attachment cross-section derived by \cite{Letaw}, based on the lab measurements of \cite{Crawford} and \cite{Wilson}, that is required to explain experimental data.}  
\label{fig:disk}
\end{figure}

We begin with the analysis of the  disk-like model. We consider a nominal halo size of $z_h=1$~kpc and a diffusion coefficient normalization of D$_0=5 \times 10^{27}$~cm$^{2}/$sec  that recovers the observed sub-Fe/Fe ratio,  and its comparison with \cite{Letaw}. The rational of using the values that fit  the sub-Fe/Fe ratio and not the B/C ratio is because the EC isotopes are much closer to Iron and the other isotopes that we consider here. 
Fig.\ \ref{fig:disk} provides a contour plot of the $\chi^2$ fit between model and observations, for the two parameters in the attachment process formula, $N$ and $\mu$.

Fig.\ \ref{fig:disk} depicts the $\chi^2$ fit to the datasets, when separately fitting the $^{51}$V/$^{51}$Cr data, the $^{49}$Ti/$^{49}$V, and fitting them together. One can easily see that there is an inconsistency. While the observations for $^{51}$V/$^{51}$Cr require a high normalization factor and a low power-law index, the observations for $^{49}$Ti/$^{49}$V require a lower normalization factor and a higher power-law index. This inconsistency between the two observations was already demonstrated in all previous works \citep{Niebur,Niebur2003,Jones,WebberSoutoulcross}.

Despite this inconsistency, the optimal power-law index for the combined $\chi^2$ for both data sets,  $\mu=2.1 \pm 0.7$,  is in agreement with \cite{Letaw}'s results, $\mu=1.8 \pm 0.1$. Even for each separate set of isotope ratio measurements, $\mu=1.8$ is inside the respective $2\sigma$ region. We note, however, that the significance contours denote only the statistical uncertainties, but not the unknown systematic errors that should exist given the uncertainty concerning the cross-section .  

In addition to the the fact that the normalization cross-section required to explain the two observations are inconsistent with each other, they are also inconsistent with \cite{Letaw} who require a somewhat larger attachment cross-section. For the disk-like model, the $^{49}$Ti/$^{49}$V data require a normalization factor of  $N=(5.6 \pm 0.8) \times 10^{-5}$mb, the $^{51}$V/$^{51}$Cr data require  $N=(8.2 \pm 0.7) \times 10^{-5}$mb, while \cite{Letaw} finds $N=(1.2 \pm 0.2) \times 10^{-4}$mb,  which is about a factor of 1.5 higher than our result.

\begin{figure}
\centerline{\includegraphics[width=3.0in]{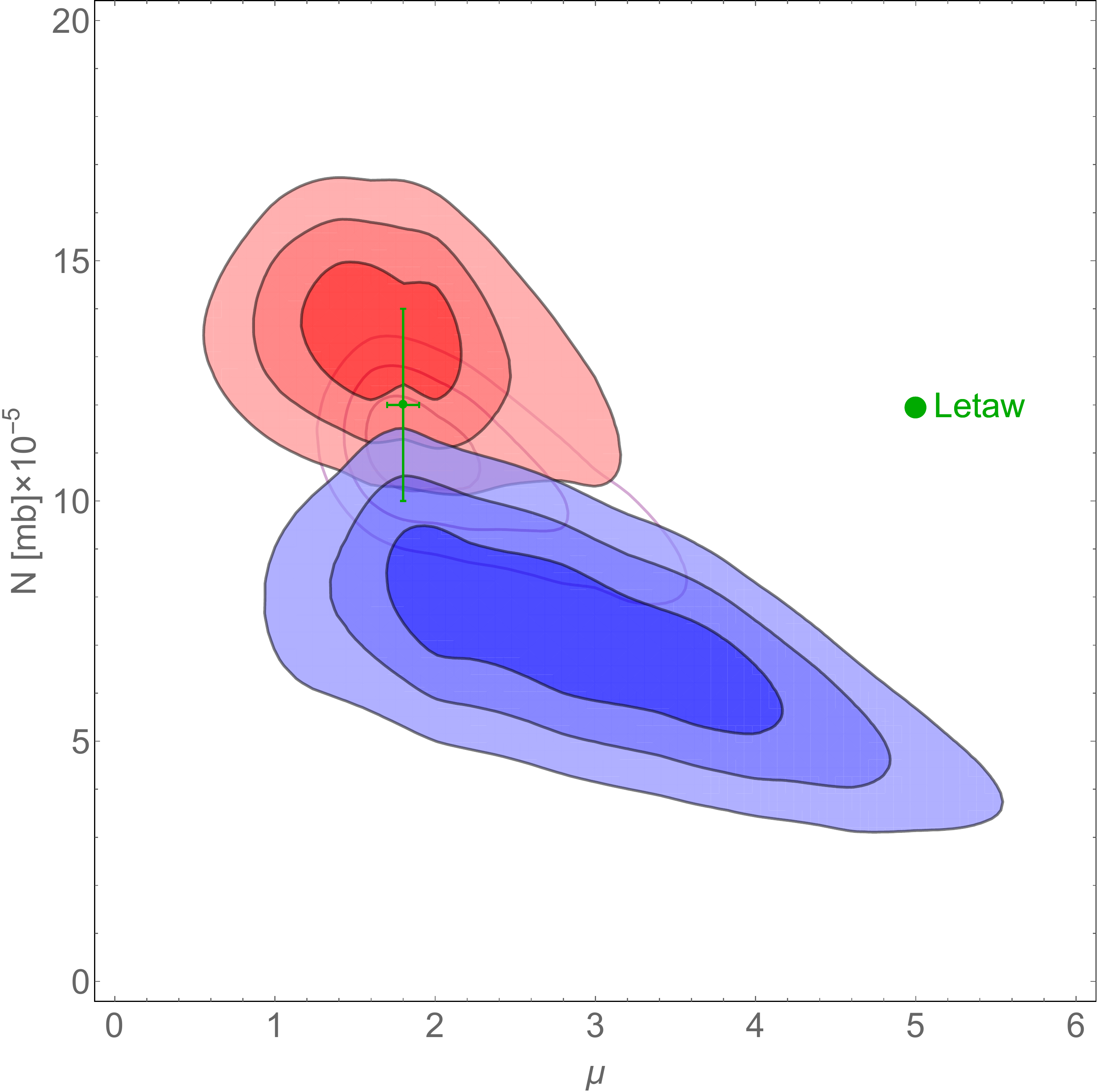}}
\caption{A contour plot of $\chi^2$ for a disk-like model with halo size of $z_h=3$~kpc and diffusion coefficient of D$_0=1.5 \times 10^{28}$~cm$^{2}/$sec. Similar to fig.\ \ref{fig:disk}, the red contours correspond to the $^{51}$V/$^{51}$Cr fit, the blue contours correspond to the $^{49}$Ti/$^{49}$V fit and the dashed purple lines correspond to the combined $\chi^2$ calculation. Note the discrepancy between the two observations remains the same as in fig.\ \ref{fig:disk}.}
\label{fig:LargeDisk}
\end{figure}

Next, we proceed to check whether the inconsistency between the two sets of observations and the inconsistency in the normalization of the cross-section  between our results and those of \cite{Letaw} are an outcome of the model parameters or whether these inconsistencies remains for all disk-like models. Fig.\ \ref{fig:LargeDisk} depicts the $\chi^2$ fit for a disk-like model with halo size of $z_h=3$~kpc and diffusion coefficient of D$_0=1.5 \times 10^{28}$~cm$^{2}/$sec.

Evidently, the discrepancy between the two sets of observations remains for different halo sizes, however, a larger halo can remove the discrepancy between the required attachment cross-section to fit the lab measurements and the average of the cross-section's normalization factor determined from the two EC datasets. 

\subsection{The normalization dependence on $z_\mathrm{h}$ in a disk-like model}
\label{sec:DiskFit}

When varying the Galactic halo size, one has to take into account other observational constraints on the secondary to primary ratios, such as B/C or sub-Iron/Iron. In fact, imposing the sub-Iron to Iron ratio measurements imposes the linear relation 
 $D_0/z_\mathrm{h}=(5 \pm 1) \times 10^{27} ~$(cm$^{2}/$sec)~kpc$^{-1}$.  Namely, for each $z_\mathrm{h}$ there is a corresponding diffusion coefficient normalization, $D_0$. 
 
 We note again that under the disk-like models, the B/C ratio requires a different normalization for the diffusion coefficient than the sub-Iron/Iron ratio \citep{Iron, Davis, GarciaMunoz}. We choose here the diffusion normalization factor  corresponding to the sub-Iron/Iron data because the EC isotopes belong to the Iron group as well. 

Next, we  fix now the attachment cross-sections power-law index, $\mu=1.8$. This value is consistent with \cite{Letaw}'s results and our results. With these constraints on $\mu$ and $D_0$, we can now proceed to obtain the fitted attachment cross-section normalization, $N_{disk}$, as a function of $z_\mathrm{h}$. 
\begin{equation} 
N_{disk}=(8.0 \pm 0.7) \times 10^{-5} \times (z_\mathrm{h}/1~\mathrm{kpc})^{0.27 \pm 0.04}~\mathrm{mb}.
\end{equation} 

One can easily see that in a disk-like model a halo size of about $z_\mathrm{h} \approx 3$ to $5$~kpc is required in order to match \cite{Letaw}'s results.

\begin{figure}
\centerline{\includegraphics[width=3.0in]{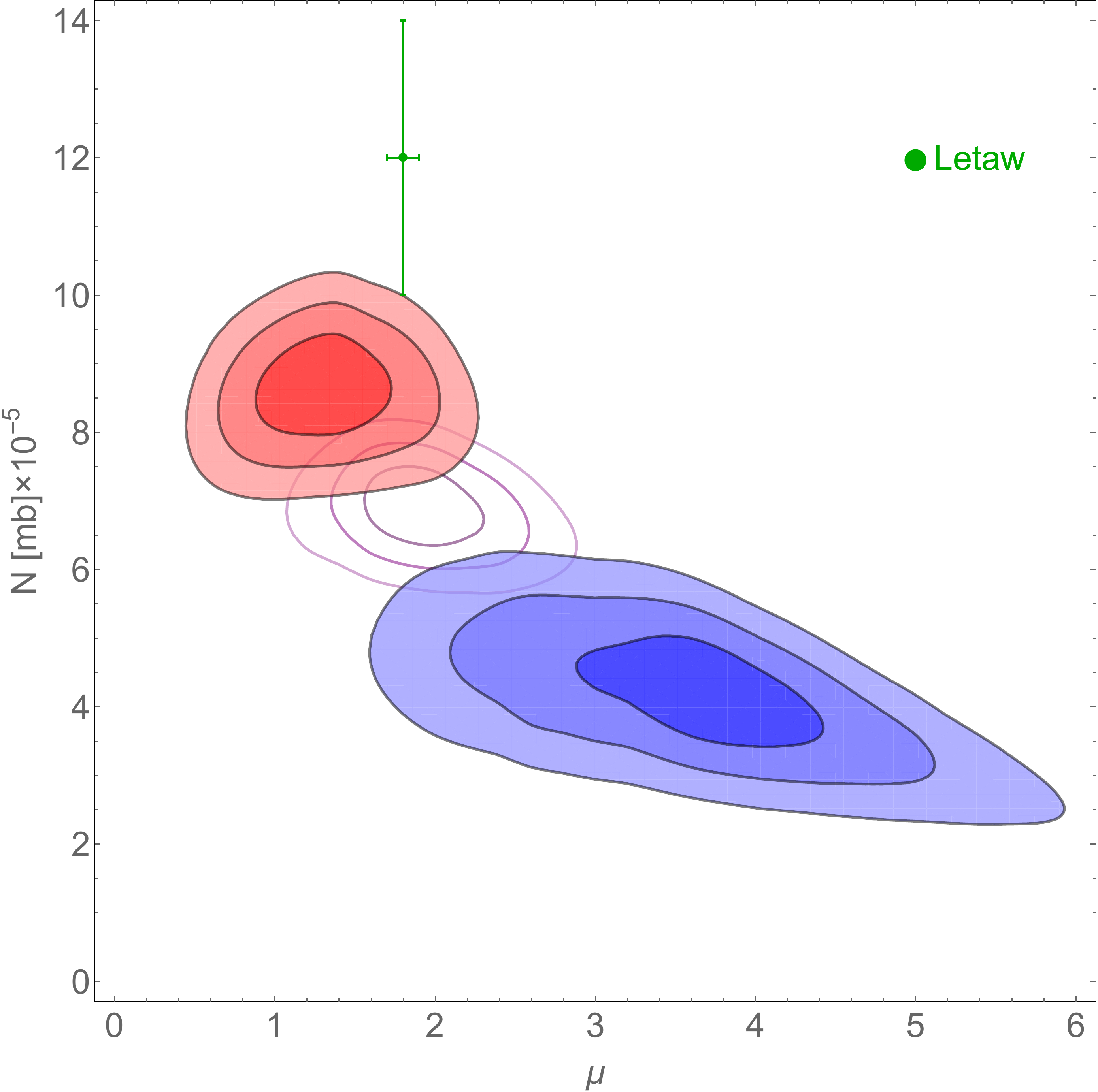}}
\caption{A contour plot of $\chi^2$ for the spiral-arms model. Similar to figs.\ \ref{fig:disk} and \ref{fig:LargeDisk}, the red contours correspond to the $^{51}$V/$^{51}$Cr fit, the blue contours correspond to the $^{49}$Ti/$^{49}$V fit and the dashed purple lines correspond to the combined $\chi^2$ calculation. Note the discrepancy between the two observations is somewhat larger but still similar in size to the  disk-like model.}
\label{fig:SA}
\end{figure}

\subsection{A spiral-arms model}
\label{sec:SA model}

We now proceed to study our nominal spiral-arm model from \cite{Iron}. In particular we are interested in finding the optimal parameters (of the attachment process formula) that recover the observations. Fig.\ \ref{fig:SA} depicts a contour map of $\chi^2$, similar to figs.\ \ref{fig:disk} and \ref{fig:LargeDisk}, but for the spiral-arms model.

As is the case in the disk-like model, the optimal power-law index for the combined $\chi^2$ calculation in the spiral-arms model is $\mu=1.9 \pm 0.4$. This value agrees with \cite{Letaw}'s result as well. Nevertheless, the inconsistency between the required cross-section normalization of the two observations remains the same as in the disk-like model and all other previous works. This suggests that by the changing the diffusion parameters or geometry one cannot resolve the discrepancy, which probably arises due to uncertainties in the cross-sections. 

 We find that the $^{49}$Ti/$^{49}$V data require a normalization of  $N=(4.3 \pm 0.7) \times 10^{-5}$mb and the $^{51}$V/$^{51}$Cr data require  $N=(8.7 \pm 0.7) \times 10^{-5}$mb. For a comparison again, \cite{Letaw} finds $N=(1.2 \pm 0.2) \times 10^{-4}$mb. 

Fig.\ \ref{fig:K} depicts the two observations, $^{49}$Ti/$^{49}$V and $^{51}$V/$^{51}$Cr, with the model prediction for the optimal parameters derived above. The shaded regions correspond to the spectrum once solar wind modulation is added\footnote{More details on the solar modulations can be found in \cite{BoverC} \S3.6. Here we use  $\phi_{max}=920$~MV and  $\phi_{min}=510$~MV which are the solar modulation values that correspond to the years of the CRIS measurements \citep{UsoskinPhi}.}.

\subsection{The normalization dependence on $z_\mathrm{h}$ and $\tau_\mathrm{arm}$ in a spiral-arms model}
\label{sec:fit}

\begin{figure}
\centerline{\includegraphics[width=3.0in]{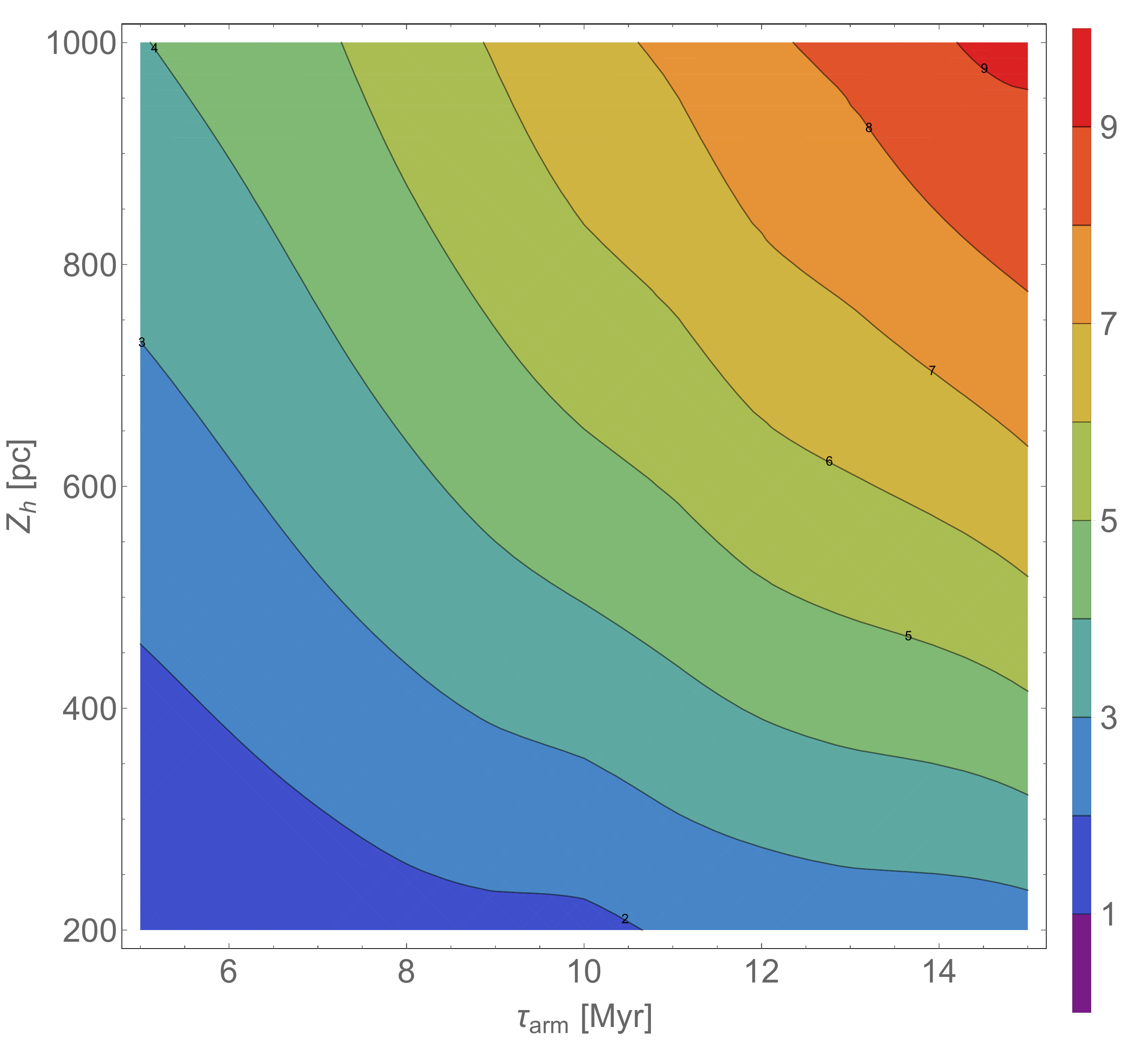}}
\caption{A contour map of the optimal $D_0$ required to fit the sub-Iron to Iron measurements, as a function of $z_\mathrm{h}$ and $\tau_\mathrm{arm}$. Evidently,  $D_0$ increases with both $z_\mathrm{h}$ and $\tau_\mathrm{arm}$.}
\label{fig:D}
\end{figure}

In a similar way to disk-like models,   observational constraints on the secondary to primary ratios, such as B/C or sub-Iron/Iron ratios, imply that the  normalization of the  diffusion coefficient, $D_0$, varies when changing the geometry of the arms and/or  the galaxy. Namely, for each pair of $z_\mathrm{h}$ and $\tau_\mathrm{arm}$, there is a corresponding value of $D_0$. While in the disk-like models we had to chose  this normalization  that will fit either the B/C ratio or the sub-Iron/Iron ratio, here in the spiral-arms model the same $D_0$ is consistent with both the B/C and sub-Iron/Iron data \citep{Iron}. Fig.\,\ref{fig:D} shows a contour map of $D_0$ as a function of $z_\mathrm{h}$ and $\tau_\mathrm{arm}$. As expected, $D_0$ increases with both $z_\mathrm{h}$ and $\tau_\mathrm{arm}$.

The next step is to fix the attachment cross-sections power-law index, $\mu=1.8$. 
This value is consistent with 
 \cite{Letaw}'s results. 

\begin{figure}
\centerline{\includegraphics[width=3.0in]{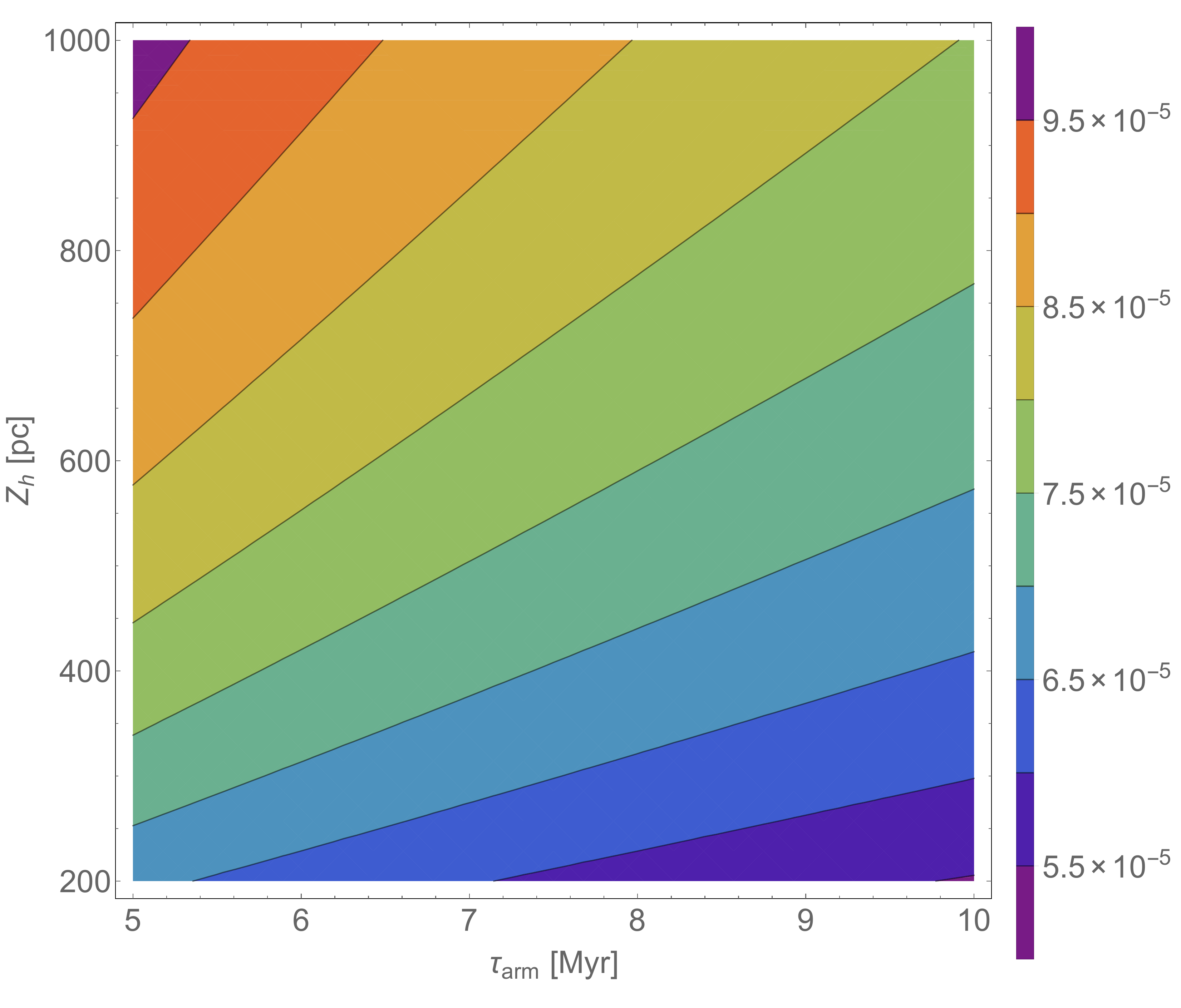}}
    \caption{A contour map of the optimal attachment cross-section normalization, $N$, required to fit the combined $\chi^2$ calculation of the two datasets, as a function of $z_\mathrm{h}$ and $\tau_\mathrm{arm}$. It is readily seen that the normalization, $N$, increases when $z_\mathrm{h}$ increases, while it decreases with $\tau_\mathrm{arm}$. Note that the rugged behavior arises from the raw data having ``Monte Carlo" noise.}
\label{fig:ACombined}
\end{figure}

With the above values of  $\mu$ and $D_0$, we can now proceed to obtain the optimal normalization of the attachment cross-section, $N$, as a function of  $z_\mathrm{h}$ and $\tau_\mathrm{arm}$. Fig.\ \ref{fig:ACombined} depicts  contour maps of $N$ for the combined $\chi^2$ fit of the two datasets as a function of $z_\mathrm{h}$ and $\tau_\mathrm{arm}$. 
One can readily see that $N$ increases with $z_\mathrm{h}$ but it decreases with  $\tau_\mathrm{arm}$. 

We can quantify  better the required normalization by using the form:
\begin{eqnarray}
N_{SA}&=&(7.98 \pm 0.02) \times 10^{-5} \nonumber \\
&& \times(\tau_\mathrm{arm}/10~\mathrm{Myr})^{-0.278 \pm 0.008} \nonumber \\
&&  \times (z_\mathrm{h}/1~\mathrm{kpc})^{0.236 \pm 0.007}~\mathrm{mb}.
\end{eqnarray}

\begin{figure*}
\centerline{\includegraphics[width=3.0in]{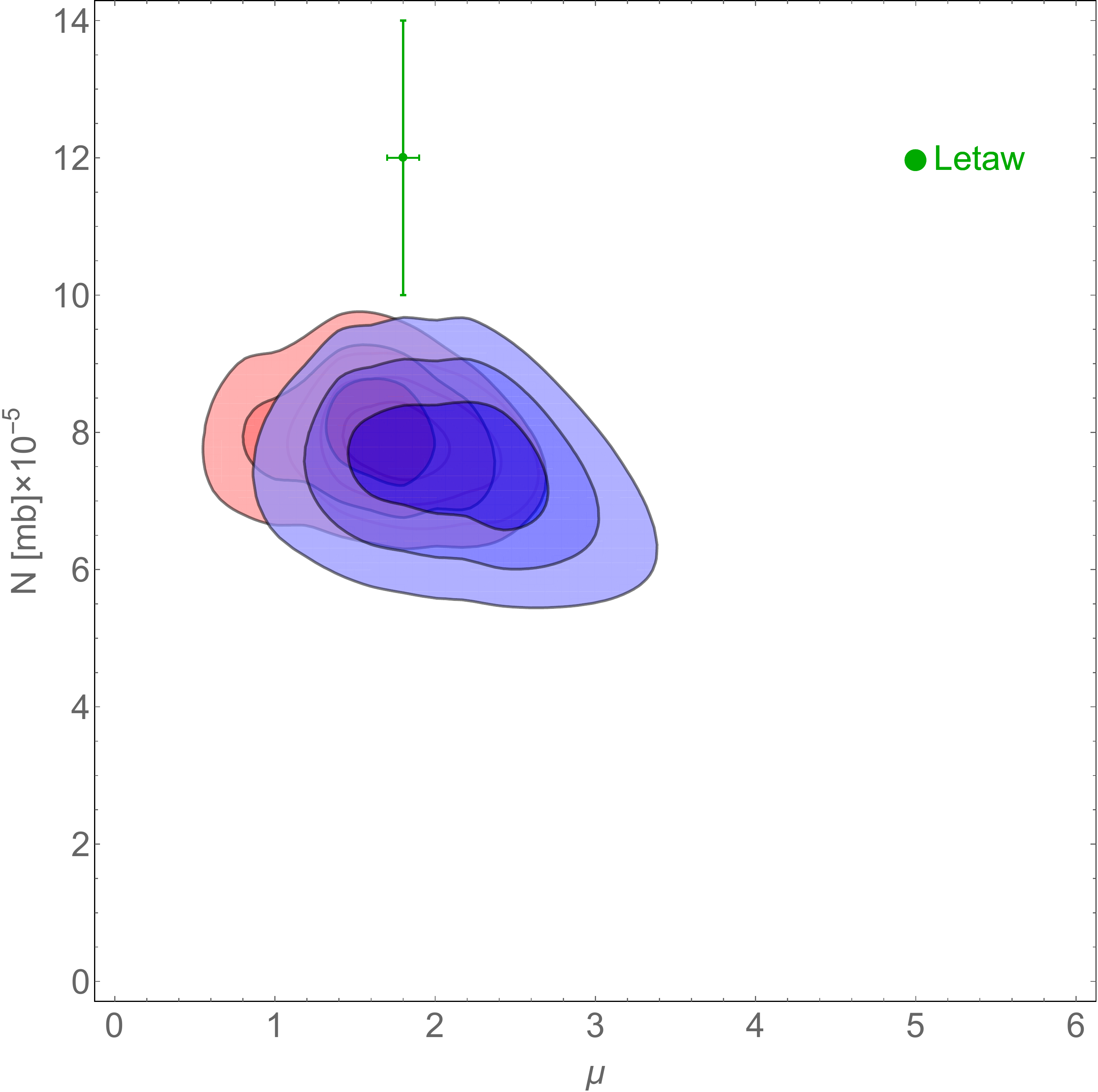} \includegraphics[width=3.0in]{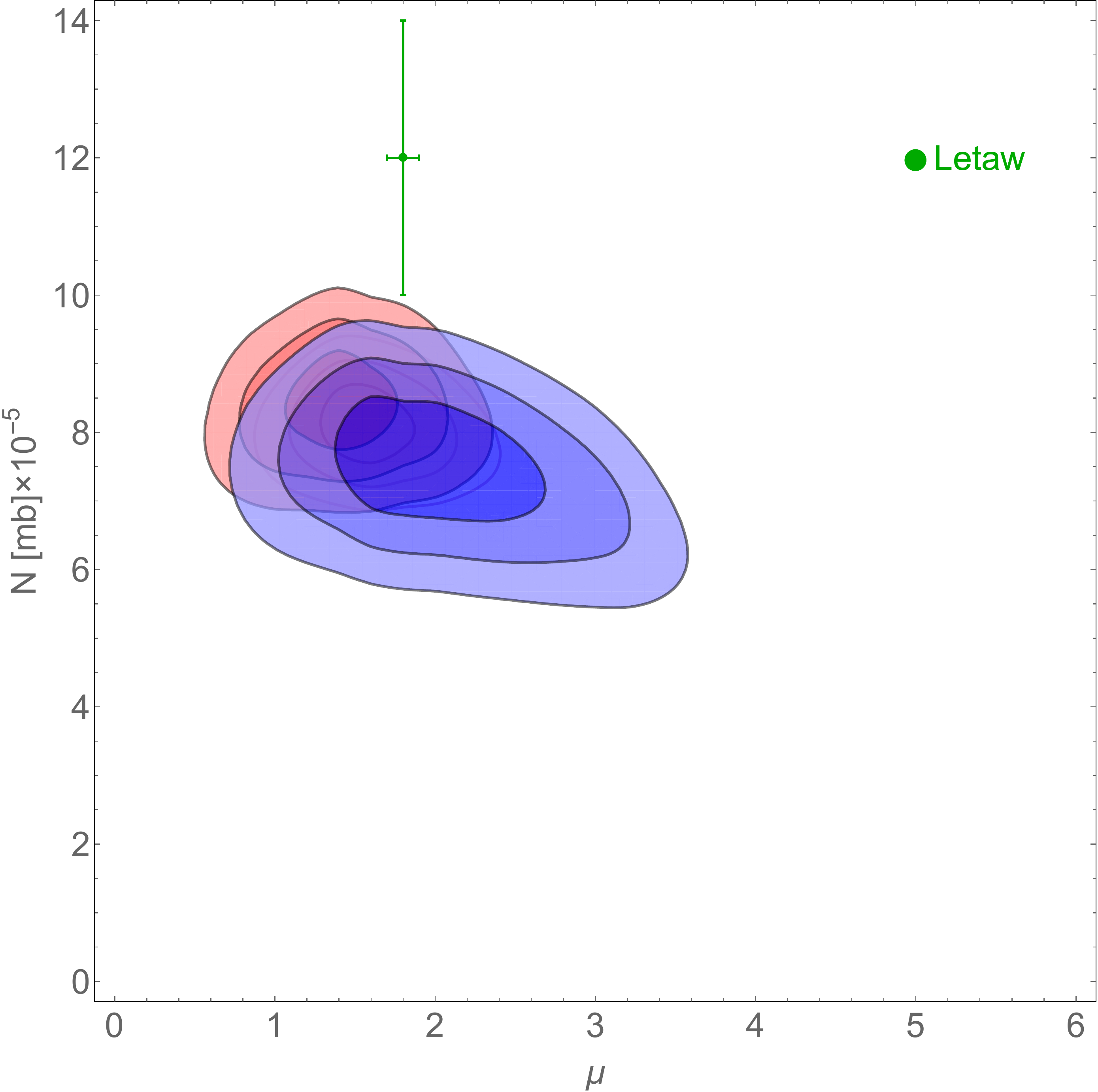}}
\caption{Contour plots of the $\chi^{2}$ fit for the $^{49}$Ti/$^{49}$V and $^{51}$V/$^{51}$Cr ratios and the combined $\chi^2$ calculation (the same colours as in fig.\ \ref{fig:disk} and fig.\ \ref{fig:SA}) for the disk-like model (left panel) and the spiral-arms model (right panel) after reducing the fragmentation cross-sections for $^{49}$Ti by 15\% and 20\% respectively.}
\label{fig:Ti}
\end{figure*}


%

\section{Discussion \& Summary}
\label{sec:discussion}

It is generally accepted that CR EC isotopes can be used to assess the importance of re-acceleration in the ISM \citep{StrongReview}.  Nonetheless, a comparison between model predictions and measurements of $^{51}$V/$^{51}$Cr and $^{49}$Ti/$^{49}$V gave inconsistent results, generally interpreted as arising from uncertainties in the nuclear spallation cross-sections. \cite{Niebur2003} have shown that the typical 10-20\% uncertainty in the fragmentation cross-sections \citep{WebberSoutoulcross} can explain away the discrepancy between the observations of the two isotopes. Specifically, they found that a reduction of  the fragmentation cross-sections of $^{49}$Ti by 15\% was sufficient to resolve the discrepancy. Previous analyses, however, considered axisymmetric models in which the CR source distribution is relatively smooth.  
 
More recently, we developed a fully 3D CR diffusion model which not only considers that most CR acceleration takes place in the vicinity of spiral arms, but also that these arms are dynamic \citep{BoverC}. One very important aspect of this model is that the path length distribution (PLD) is different from the one found in standard disk-like models. In the latter, the PLD is typically close to being exponential. However, if most CRs arrive from a distance, such as from a spiral-arm, then the PLD will exhibit a paucity of small path lengths (compare fig.\,4 to fig.\,6 in \citealt{BoverC}). It was therefore our goal to see whether a more realistic distribution of CR sources could alleviate the discrepancy between the model predictions and the measurements of $^{51}$V/$^{51}$Cr and $^{49}$Ti/$^{49}$V ratios. 

In this work, we studied the EC isotopes using the observations of $^{51}$V/$^{51}$Cr and $^{49}$Ti/$^{49}$V ratios, as well as an empirical fit to the electron attachment cross-section of propagating nuclei. This fit is based on the results of \cite{Letaw}, who derived the attachment and stripping cross-section using experimental data from \cite{Wilson} and \cite{Crawford}. They measured the time scales of both processes which are much longer than the EC decay timescale, thus, we can neglect the stripping process and assume that when an EC isotope attaches electron, it will decay immediately. \cite{Letaw} also showed that the attachment cross-section has an approximate power-law dependance on the energy and on $Z$, with respective power indices of $\mu=1.8 \pm 0.1$ and $\nu=4.5 \pm 0.1$, and a normalization of $N=(1.2 \pm 0.2) \times 10^{-4}~$mb (for $E=500~$MeV and $Z=1$).

We first found that the EC ratios in standard disk-like models are not only sensitive to the EC rates but also modestly sensitive to the halo size. Specifically, the required cross-section in disk-like models is $\sigma_{a}(E,Z)=N(z_\mathrm{h}) \times Z^{4.5} \times {(E/\mathrm{500\,MeV})^{-1.8}}$, with the normalization roughly given by $N_{disk}(z_\mathrm{h})=8 \times 10^{-5} ~$mb$ \times (z_\mathrm{h}/1~$kpc$)^{0.27}$. 

We then found that EC in spiral-arms models can also constrain the geometry of the galactic arms in addition to the halo size. The required cross-section also depends on the time since last spiral arm passage. Its normalization should satisfy $N_{SA}(z_\mathrm{h},\tau_\mathrm{arm})=7.98 \times 10^{-5} ~$mb$ \times (\tau_\mathrm{arm}/10~$Myr$)^{-0.278} \times (z_\mathrm{h}/1~$kpc$)^{0.236}$.

However, even with the added spiral arms one cannot alleviate the discrepancy between the $^{51}$V/$^{51}$Cr and $^{49}$Ti/$^{49}$V measurements. This  strengthens the claim that  this discrepancy  is due to the uncertainty in the spallation cross-sections. Thus,  improved spallation cross-sections are required
in order to use the  EC CRs  to constrain geometric properties of the diffusion models.


~\\

\section*{Acknowledgements} 
This work is supported by an Advanced ERC grant  (TP), the Israel Science Foundation (grant no.\ 1423/15, NS) and by the I-CORE Program of the Planning and Budgeting Committee and The Israel Science Foundation (1829/12).

\def\jcap{J.\ Cos.\ Astropart.\ Phys.}
\def\na{N.\ Astron.}

\bibliography{EC}

\begin{thebibliography}{}

\bibitem[\protect\citename{{Adriani}, }2009]{Adriani}
{Adriani}, O. et~al. 2009.
\newblock {An anomalous positron abundance in cosmic rays with energies
  1.5-100GeV}.
\newblock {\em \nat}, {\bf 458}(Apr.), 607--609.

\bibitem[\protect\citename{{Aharonian} {\em et~al.\ }\relax, }1995]{Pulsar3}
{Aharonian}, F.~A., {Atoyan}, A.~M., \& {Voelk}, H.~J. 1995.
\newblock {High energy electrons and positrons in cosmic rays as an indicator
  of the existence of a nearby cosmic tevatron}.
\newblock {\em \aap}, {\bf 294}(Feb.), L41--L44.

\bibitem[\protect\citename{{Benyamin} {\em et~al.\ }\relax, }2014]{BoverC}
{Benyamin}, D., {Nakar}, E., {Piran}, T., \& {Shaviv}, N.~J. 2014.
\newblock {Recovering the Observed B/C Ratio in a Dynamic Spiral-armed Cosmic
  Ray Model}.
\newblock {\em \apj}, {\bf 782}(Feb.), 34.

\bibitem[\protect\citename{{Benyamin} {\em et~al.\ }\relax, }2016]{Iron}
{Benyamin}, D., {Nakar}, E., {Piran}, T., \& {Shaviv}, N.~J. 2016.
\newblock {The B/C and Sub-iron/Iron Cosmic Ray Ratios--Further Evidence in
  Favor of the Spiral-Arm Diffusion Model}.
\newblock {\em \apj}, {\bf 826}(July), 47.

\bibitem[\protect\citename{{Bergstr{\"o}m} {\em et~al.\ }\relax,
  }2008]{DarkMetter1}
{Bergstr{\"o}m}, L., {Bringmann}, T., \& {Edsj{\"o}}, J. 2008.
\newblock {New positron spectral features from supersymmetric dark matter: A
  way to explain the PAMELA data?}
\newblock {\em \prd}, {\bf 78}(10), 103520--+.

\bibitem[\protect\citename{{Cesarsky}, }1980]{CRreview}
{Cesarsky}, C.~J. 1980.
\newblock {Cosmic-ray confinement in the galaxy}.
\newblock {\em \araa}, {\bf 18}, 289--319.

\bibitem[\protect\citename{{Chi} {\em et~al.\ }\relax, }1996]{Pulsar2}
{Chi}, X., {Cheng}, K.~S., \& {Young}, E.~C.~M. 1996.
\newblock {Pulsar Wind Origin of Cosmic Ray Positrons}.
\newblock {\em \apjl}, {\bf 459}(Mar.), L83+.

\bibitem[\protect\citename{Connell, }1999]{connell1999ulysses}
Connell, James. 1999.
\newblock Ulysses HET measurements of electron-capture secondary isotopes:
  Testing the role of cosmic ray reacceleration.
\newblock {\em Page ~33 of:} {\em International Cosmic Ray Conference},  vol.
  3.

\bibitem[\protect\citename{{Crawford}, }1979]{Crawford}
{Crawford}, H.~J. 1979.
\newblock {\em {Single electron attachment and stripping cross sections for
  relativistic heavy ions}}.
\newblock Ph.D. thesis, California Univ., Berkeley.

\bibitem[\protect\citename{{Davis} {\em et~al.\ }\relax, }2000]{Davis}
{Davis}, A.~J., {Mewaldt}, R.~A., {Binns}, W.~R., {Christian}, E.~R.,
  {Cummings}, A.~C., {George}, J.~S., {Hink}, P.~L., {Leske}, R.~A., {von
  Rosenvinge}, T.~T., {Wiedenbeck}, M.~E., \& {Yanasak}, N.~E. 2000 (Sept.).
\newblock {On the low energy decrease in galactic cosmic ray secondary/primary
  ratios}.
\newblock {\em Pages  421--424 of:} {Mewaldt}, R.~A., {Jokipii}, J.~R., {Lee},
  M.~A., {M{\"o}bius}, E., \& {Zurbuchen}, T.~H. (eds), {\em Acceleration and
  Transport of Energetic Particles Observed in the Heliosphere}.
\newblock American Institute of Physics Conference Series, vol. 528.

\bibitem[\protect\citename{{di Bernardo} {\em et~al.\ }\relax, }2010]{Dragon}
{di Bernardo}, G., {Evoli}, C., {Gaggero}, D., {Grasso}, D., \& {Maccione}, L.
  2010.
\newblock {Unified interpretation of cosmic ray nuclei and antiproton recent
  measurements}.
\newblock {\em Astroparticle Physics}, {\bf 34}(Dec.), 274--283.

\bibitem[\protect\citename{DuVernois {\em et~al.\ }\relax,
  }1996]{duvernois1996interstellar}
DuVernois, MA, Simpson, JA, \& Thayer, MR. 1996.
\newblock Interstellar propagation of cosmic rays: analysis of the ULYSSES
  primary and secondary elemental abundances.
\newblock {\em Astronomy and Astrophysics}, {\bf 316}, 555--563.

\bibitem[\protect\citename{{Garcia-Munoz} {\em et~al.\ }\relax,
  }1987]{GarciaMunoz}
{Garcia-Munoz}, M., {Simpson}, J.~A., {Guzik}, T.~G., {Wefel}, J.~P., \&
  {Margolis}, S.~H. 1987.
\newblock {Cosmic-ray propagation in the Galaxy and in the heliosphere - The
  path-length distribution at low energy}.
\newblock {\em \apjs}, {\bf 64}(May), 269--304.

\bibitem[\protect\citename{{Harding} \& {Ramaty}, }1987]{Pulsar1}
{Harding}, A.~K., \& {Ramaty}, R. 1987.
\newblock {The Pulsar Contribution to Galactic Cosmic Ray Positrons}.
\newblock {\em Pages  92--+ of:} {\em International Cosmic Ray Conference}.
\newblock International Cosmic Ray Conference, vol. 2.

\bibitem[\protect\citename{{Hooper} {\em et~al.\ }\relax, }2009]{Pulsar4}
{Hooper}, D., {Blasi}, P., \& {Dario Serpico}, P. 2009.
\newblock {Pulsars as the sources of high energy cosmic ray positrons}.
\newblock {\em \jcap}, {\bf 1}(Jan.), 25--+.

\bibitem[\protect\citename{{Ibarra} \& {Tran}, }2008]{DarkMetter2}
{Ibarra}, A., \& {Tran}, D. 2008.
\newblock {Antimatter signatures of gravitino dark matter decay}.
\newblock {\em \jcap}, {\bf 7}(July), 2--+.

\bibitem[\protect\citename{{Jones} {\em et~al.\ }\relax, }2001]{Jones}
{Jones}, F.~C., {Lukasiak}, A., {Ptuskin}, V.~S., \& {Webber}, W.~R. 2001.
\newblock {K-Capture cosmic ray secondaries and reacceleration}.
\newblock {\em International Cosmic Ray Conference}, {\bf 5}(Aug.), 1844.

\bibitem[\protect\citename{{Letaw} {\em et~al.\ }\relax, }1985]{Letaw}
{Letaw}, J.~R., {Adams}, Jr., J.~H., {Silberberg}, R., \& {Tsao}, C.~H. 1985.
\newblock {Electron capture decay of cosmic rays}.
\newblock {\em \apss}, {\bf 114}(Sept.), 365--379.

\bibitem[\protect\citename{{Niebur} {\em et~al.\ }\relax, }2000]{ACE}
{Niebur}, S.~M., {Binns}, W.~R., {Christian}, E.~R., {Cummings}, A.~C.,
  {George}, J.~S., {Hink}, P.~L., {Israel}, M.~H., {Klarmann}, J., {Leske},
  R.~A., {Lijowski}, M., {Mewaldt}, R.~A., {Stone}, E.~C., {von Rosenvinge},
  T.~T., {Wiedenbeck}, M.~E., \& {Yanasak}, N.~E. 2000 (Sept.).
\newblock {Secondary electron-capture-decay isotopes and implications for the
  propagation of galactic cosmic rays}.
\newblock {\em Pages  406--409 of:} {Mewaldt}, R.~A., {Jokipii}, J.~R., {Lee},
  M.~A., {M{\"o}bius}, E., \& {Zurbuchen}, T.~H. (eds), {\em Acceleration and
  Transport of Energetic Particles Observed in the Heliosphere}.
\newblock American Institute of Physics Conference Series, vol. 528.

\bibitem[\protect\citename{{Niebur} {\em et~al.\ }\relax, }2001]{Niebur}
{Niebur}, S.~M., {Binns}, W.~R., {Christian}, E.~R., {Cummings}, A.~C., {de
  Nolfo}, G.~A., {George}, J.~S., {Hink}, P.~L., {Israel}, M.~H., {Leske},
  R.~A., {Mewaldt}, R.~A., {Stone}, E.~C., {von Rosenvinge}, T.~T.,
  {Wiedenbeck}, M.~E., \& {Yanasak}, N.~E. 2001.
\newblock {CRIS measurements of electron-capture decay isotopes: 37Ar, 44Ti,
  49V, 51Cr, 55Fe, and 57Co}.
\newblock {\em International Cosmic Ray Conference}, {\bf 5}(Aug.), 1675.

\bibitem[\protect\citename{{Niebur} {\em et~al.\ }\relax, }2003]{Niebur2003}
{Niebur}, S.~M., {Scott}, L.~M., {Wiedenbeck}, M.~E., {Binns}, W.~R.,
  {Christian}, E.~R., {Cummings}, A.~C., {Davis}, A.~J., {George}, J.~S.,
  {Hink}, P.~L., {Israel}, M.~H., {Leske}, R.~A., {Mewaldt}, R.~A., {Stone},
  E.~C., {von Rosenvinge}, T.~T., \& {Yanasak}, N.~E. 2003.
\newblock {Cosmic ray energy loss in the heliosphere: Direct evidence from
  electron-capture-decay secondary isotopes}.
\newblock {\em Journal of Geophysical Research (Space Physics)}, {\bf
  108}(Oct.), 8033.

\bibitem[\protect\citename{Oliva {\em et~al.\ }\relax, }2013]{AMS-preliminary}
Oliva, A, Collaboration, AMS, {\em et~al.\ }\relax. 2013.
\newblock Precision measurement of the cosmic ray boron-to-carbon ration with
  AMS.
\newblock {\em Proceedings 33rd ICRC, Rio de Janeiro}.

\bibitem[\protect\citename{{Profumo}, }2008]{Pulsar5}
{Profumo}, S. 2008.
\newblock {Dissecting cosmic-ray electron-positron data with Occam's Razor: the
  role of known Pulsars}.
\newblock {\em ArXiv e-prints}, Dec.

\bibitem[\protect\citename{{Shaviv} {\em et~al.\ }\relax, }2009]{Pamela}
{Shaviv}, N.~J., {Nakar}, E., \& {Piran}, T. 2009.
\newblock {Inhomogeneity in Cosmic Ray Sources as the Origin of the Electron
  Spectrum and the PAMELA Anomaly}.
\newblock {\em Physical Review Letters}, {\bf 103}(11), 111302--+.

\bibitem[\protect\citename{{Strong} \& {Moskalenko}, }1998]{StrongNucleons}
{Strong}, A.~W., \& {Moskalenko}, I.~V. 1998.
\newblock {Propagation of Cosmic-Ray Nucleons in the Galaxy}.
\newblock {\em \apj}, {\bf 509}(Dec.), 212--228.

\bibitem[\protect\citename{{Strong} {\em et~al.\ }\relax, }2007]{StrongReview}
{Strong}, A.~W., {Moskalenko}, I.~V., \& {Ptuskin}, V.~S. 2007.
\newblock {Cosmic-Ray Propagation and Interactions in the Galaxy}.
\newblock {\em Annual Review of Nuclear and Particle Science}, {\bf 57}(Nov.),
  285--327.

\bibitem[\protect\citename{Usoskin {\em et~al.\ }\relax, }2011]{UsoskinPhi}
Usoskin, I.G., Kovaltsov, GA, \& Bazilevskaya, GA. 2011.
\newblock Solar modulation parameter for cosmic rays since 1936 reconstructed
  from ground-based neutron monitors and ionization chambers.
\newblock {\em J. Geophys. Res}, {\bf 116}, A02104.

\bibitem[\protect\citename{{Webber} {\em et~al.\ }\relax,
  }2003]{WebberSoutoulcross}
{Webber}, W.~R., {Soutoul}, A., {Kish}, J.~C., \& {Rockstroh}, J.~M. 2003.
\newblock {Updated Formula for Calculating Partial Cross Sections for Nuclear
  Reactions of Nuclei with $Z<=28$ and $E>150$ MeV Nucleon$^{-1}$ in Hydrogen
  Targets}.
\newblock {\em \apjs}, {\bf 144}(Jan.), 153--167.

\bibitem[\protect\citename{{Wilson}, }1978]{Wilson}
{Wilson}, L.~W. 1978.
\newblock {\em {The nuclear and atomic physics governing changes in the
  composition of relativistic cosmic rays}}.
\newblock Ph.D. thesis, California Univ., Berkeley.

\end{thebibliography}

\end{document}